\documentclass[twoside,8pt]{article}
\usepackage{epsfig}
\usepackage{amsmath,amsfonts}
\usepackage{color}
\usepackage{xcolor}

\newcommand{\be}{\begin{equation}}
\newcommand{\ee}{\end{equation}}
\newcommand{\bea}{\begin{eqnarray}}
\newcommand{\eea}{\end{eqnarray}}

\begin{document}

\title{Constraints on the equation of state from the stability condition  of neutron stars}

\author{
P.S.  Koliogiannis$^1$ and Ch.C. Moustakidis$^{1,2}$ (moustaki@auth.gr)\\
$^1$Department of Theoretical Physics, Aristotle University of
Thessaloniki,\\  54124 Thessaloniki, Greece\\
$^{2}$Theoretical Astrophysics, IAAT, Eberhard-Karls University of Tuebingen,\\
72076 Tuebingen, Germany  }

\maketitle

\begin{abstract}
The  stellar equilibrium and collapse, including mainly white dwarfs, neutron stars and supper massive stars,  is an interplay between general relativistic effects and the equation of state of nuclear matter. In the present work, we use  the Chandrasekhar criterion of stellar instability by employing a large number of realistic equations of state (EoS) of neutron star matter. We mainly focus on the critical point of transition from stable to unstable configuration. This point corresponds to the maximum neutron star mass configuration. We calculate, in each case, the resulting compactness parameter, $\beta=GM/c^2R$, and the corresponding effective adiabatic index, $\gamma_{\rm cr}$. The role of the trial function $\xi(r)$ is presented and discussed in details. We found that it holds a model-independent relation  between $\gamma_{\rm cr}$ and $\beta$. This statement is strongly supported by the large number of EoS and it is also corroborated by using  analytical solutions of the Einstein's field equations.  In addition, we present and discuss  the relation between the maximum rotation rate and the adiabatic index close to the instability limit. Accurate observational measurements of the upper bound of the neutron star mass and the corresponding radius, in connection with the present predictions, may help to impose constraints on the high density part of the neutron star equation of state.
\\
\\
%
\\
PACS number(s): 97.60.Jd; 26.60 Kp; 04.40.Dg.

\end{abstract}

\section{Introduction}
The discovery of the inspiral and coalescence of a binary neutron star system (GW170817), by the Laser Interferometer Gravitational-wave Observatory (LIGO, VIRGO) (on 2017, August 17), open a new window to explore the neutron star equation of state at high densities~\cite{Abbott-017,Ligo-017}. In particular, just after the mentioned discovery, a significant effort was put in constraining  the upper as well as the lower limit of the maximum neutron star mass and the corresponding radius. In any case, one of the main ingredient is the compactness of the neutron star, which is expected to play important role
in the stability and dynamics processes of  neutron stars. It is well known that the maximum mass, which corresponds to the most compact configuration, is the border between the stable-unstable configuration. Very useful and robust information can be gained by studying  this extreme case.

The stability of  relativistic stars has been studied extensively in the past~\cite{Chandrasekhar-64,Chandrasekhar-64b,Weinberg-72,Harrison-65,Zeldovich-78,Shapiro-83,Glendenning-2000,Haensel-07,Friedman-013}  while  various approaches have been used  in order to treat  this problem~\cite{Bardeen-66}. In particular,  firstly one can solve the Tolman-Oppenheimer-Volkoff (TOV)~\cite{Tolman-39,Oppenheimer-39}  equations
(which provide the   equilibrium configuration) for either  numerically derived  equation of state or trying to find analytical solutions. In any case, both of the solutions lead to an infinite number of configurations. Secondly, one possibility is the use of the  criterion of  Chandrasekhar~\cite{Chandrasekhar-64,Chandrasekhar-64b}  in order to identify, in each case, the stable configurations as well as  the interface between stable and unstable configuration.

It is worth to pointing out that in order to extract a solution with physical interest, one have to solve the Einstein's field equations using a realistic equation of state of the fluid interior. However, there are a few analytical solutions with  physical interest which may help to introduce and to establish  some universal approximations.


Moreover, at a given density, there is an important parameter that is called adiabatic index and in particular, characterizes  the stiffness of the equation of state~\cite{Harrison-65,Haensel-07,Misner-73,Bludman-73a,Bludman-73b,Ipser-70,Glass-83,Lindblom-83,Hiscock-83,Gaertig-09}. The instability criterion of Chandrasekhar~\cite{Chandrasekhar-64,Chandrasekhar-64b}, strongly depends on this parameter (adiabatic index). One of the main motivation of the present is to examine the   possibility to impose constraints on the realistic neutron star equations of state via  the instability condition of Chandrasekhar.

In particular, we employ an extended group of realistic equations of state based on various theoretical nuclear models. The abbreviated names of these equations of state are: MDI~\cite{Prakash-97,Moustakidis-08}, NLD~\cite{Gaitanos-013,Gaitanos-015}, HHJ~\cite{Heiselberg -2000}, Ska, SkI4~\cite{Chabanat-97,Farine-97}, HLPS~\cite{Hebeler-013}, SCVBB~\cite{Sharma-015}, BS~\cite{Balberg-2000}, BGP~\cite{Bowers-75},
W~\cite{Walecka-74}, DH~\cite{Douchin-01}, BL~\cite{Bombaci-018}, WFF1,WFF2~\cite{Wiringa-88}, APR~\cite{Akmal-98} and PS~\cite{Pandha-75}. All of them satisfy, at least marginally, the observed limit of $M = 1.97\pm 0.04 \ M_{\odot}$ (PRS J1614-2230~\cite{Demorest-10}) and $M = 2.01 \pm 0.04 \ M_{\odot}$ (PSRJ0348+0432~\cite{Antoniadis-13}).
Actually, at the moment, the most robust constraints on the neutron star equations of state are based on the measurements of the lower bound of the maximum neutron star mass. Strictly speaking, the suggested equations of state which do not reproduce the higher measurement of neutron star mass, must be excluded.

It is well known also that the rapidly rotating neutron stars can be used in order to determine the equation of state (see Ref.~\cite{Haensel-07} and reference therein). In particular, the maximum rotating frequency $f_{\rm max}$ (Keplerian frequency) depends both on the gravitational mass $M_{\rm max}$ and the EoS. Until this moment, the fastest known pulsar, PSR J1748-244ad, is rotating with frequency $716$ Hz~\cite{Hessels-06}. While the theoretical predicted values for $f_{\rm max}$ are much more  higher than $716$ Hz, there is a lack of neutron stars rotating faster than this value. This is an open problem and obviously additional theoretical assumptions must be in order to solve it.

In the present work we concentrate our study  on the dependence  of the effective critical adiabatic index  on the compactness of neutron star for each equation of state. We mainly focus on the interface between stable and unstable configuration which corresponds to the maximum mass configuration. This region is very important since it is directly related with the high density part of the neutron star equation of state. This issue still remains an open problem. Moreover, we propose an additional method  to constraint the equations of state by the help of accurate measurements of the maximum neutron star mass  and/or compactness. Finally, we make an effort to relate the maximum rotating frequency $f_{\rm max}$ with the critical adiabatic index and the bulk properties corresponding to the maximum mass configuration of a non-rotating (static) neutron star (including the maximum mass $M_{\rm max}^{\rm stat}$, the corresponding radius $R_{\rm max}^{\rm stat}$ and the compactness parameter $\beta_{\rm max}^{\rm stat}$) and to indicate how observational measurements of high rotating neutron stars may impose constraints on the EoS.

The article is organized as follows. In Section~2 we present the TOV equations, the Chandrasekhar's instability criterion, the definition of the relevant adiabatic indices and we briefly present four relevant  analytical solutions of the TOV equations.  In section~3 we briefly discuss the maximum rotating frequency in connection with the maximum mass configuration.
The results are presented and discussed in Section~4 and Section~5 contains the concluding remarks of the study. The appendix contains  relevant
analytical approximations for the critical adiabatic index.

\section{The stability criterion and the adiabatic indices  }
The starting point for determining  the mechanical equilibrium of  neutron star matter is the well known
Tolman-Oppenheimer-Volkoff (TOV) equations~\cite{Shapiro-83,Glendenning-2000,Tolman-39,Oppenheimer-39}. This set of differential equations  describes the structure of a neutron star.  For  a static spherical symmetric system, the metric read as follow~\cite{Shapiro-83,Glendenning-2000}
\begin{equation}
ds^2=-e^{\nu(r)}c^2dt^2+e^{\lambda(r)}dr^2+r^2\left(d\theta^2+\sin^2\theta d\phi^2\right),
\label{GRE-1}
\end{equation}
and the corresponding TOV equations take the form
\begin{equation}
\frac{dP(r)}{dr}=-\frac{G{\cal E}(r) M(r)}{c^2r^2}\left(1+\frac{P(r)}{{\cal E}(r)}\right)\left(1+\frac{4\pi P(r) r^3}{M(r)c^2}\right) \left(1-\frac{2GM(r)}{c^2r}\right)^{-1},
\label{TOV-1}
\end{equation}
\begin{equation}
\frac{dM(r)}{dr}=\frac{4\pi r^2}{c^2}{\cal E}(r).
\label{TOV-2}
\end{equation}
By introducing a realistic  EoS for neutron star (e.g. a dependence on the form $P=P({\cal E})$)  we solve numerically the TOV equations. This EoS provides the relation between pressure and density of  neutron star matter. Of course, one can try to find out analytical solutions of the TOV equations. However, it is worth to pointing out that using the analytical solutions, although each of any analytical solution describes equilibrium configurations, is not sufficient to tell us if it corresponds to stable  configurations~\cite{Tolman-39}; this is the case also  for any numerical solution. Straightforwardly  speaking,  any unstable solution  is not of physical interest.

Chandrasekhar, in order to solve the instability problem, introduced a criterion for dynamical stability based on the variational method~\cite{Chandrasekhar-64}.
In the present work we will present this criterion with the help of the averaged  ($\langle \gamma \rangle $)  and the critical ($\gamma_{cr}$) adiabatic indices. To be more specific, the averaged adiabatic index is defined as~\cite{Merafina-89,Negi-2001,Moustakidis-017}
\begin{equation}
\langle \gamma \rangle=\frac{\displaystyle \int_{0}^R e^{(\lambda+3\nu)/2}\gamma(r) \frac{P}{r^2}\left(\frac{d}{dr}(r^2e^{-\nu/2}\xi(r))  \right)^2 dr  }
{ \displaystyle \int_{0}^R e^{(\lambda+3\nu)/2}\frac{P}{r^2}\left(\frac{d}{dr}(r^2e^{-\nu/2}\xi(r))  \right)^2 dr}.
\label{gamma-MV-1}
\end{equation}
The critical adiabatic index is given by
\begin{eqnarray}
\gamma_{cr}&=&\left[ -4\int_{0}^R e^{(\lambda+\nu)/2} r\left(\frac{ dP}{dr}\right)\xi^2 dr+\int_{0}^R e^{(\lambda+\nu)/2}\left(\frac{d P}{d r}  \right)^2\frac{r^2\xi^2}{P+ {\cal E}} dr \right.\nonumber\\
&-&\left.\frac{8\pi G}{c^4}\int_0^R e^{(3\lambda+\nu)/2}P (P+{\cal E}) r^2\xi^2 dr\right]\times\left[\int_{0}^R e^{(\lambda+3\nu)/2}\frac{ P}{r^2}\left(\frac{d}{dr}(r^2e^{-\nu/2}\xi)  \right)^2 dr\right]^{-1}.
\label{gri-term-gen}
\end{eqnarray}
The Chandrasekhar stability condition  leads to the inequality~\cite{Moustakidis-017}
\begin{equation}
\langle \gamma \rangle \geq \gamma_{cr}
\label{ineq-gamma}
\end{equation}
while the case $\langle \gamma \rangle = \gamma_{cr}$ corresponds to onset of the instability.  According to Eqs.~(\ref{gamma-MV-1}) and (\ref{gri-term-gen}) the averaged  and the critical adiabatic indices are functional  of the function $\xi(r)$ as well as of the compactness parameter $\beta$.  In particular, the lagrangian displacement away from equilibrium has the form $\zeta(r)=\xi(r)e^{-i\sigma t}$, where $\sigma$ is the pulsation frequency  of the oscillations. It is obvious from the lagrangian displacement that $\sigma^2$ can take both positive and negative values. To be more specific, a positive value of $\sigma^2$ corresponds to stable configuration while a negative one to unstable one~\cite{Chandrasekhar-64,Merafina-89,Kokkotas-01}.
 It is worth to pointing out that the stability condition~(\ref{ineq-gamma})  expresses a minimal and not just an external principle~\cite{Chandrasekhar-64}.
Obviously, there are  infinite numbers of  trial functions $\xi(r)$. The most frequently used are the following (where the names which mentioned in the paper have also indicated)
\begin{equation}
\xi(r)=r e^{\nu/2}, \quad ({\rm TF-1})
\label{xi-1}
\end{equation}
\begin{equation}
\xi(r)=r e^{\nu/4}, \quad ({\rm TF-2})
\label{xi-2}
\end{equation}
\begin{equation}
\xi(r)=r\left(1+a_1r^2+a_2r^4+a_3r^6\right)e^{\nu/2}, \quad ({\rm TF-3})
\label{xi-3}
\end{equation}
\begin{equation}
\xi(r)=r. \quad ({\rm TF-4})
\label{xi-4}
\end{equation}
Now, considering  an adiabatic perturbation, the adiabatic index $\gamma$, is defined as following~\cite{Chandrasekhar-64,Merafina-89}
\begin{equation}
\gamma\equiv \frac{P+{\cal E}}{P}\left(\frac{\partial P}{\partial{\cal E}}\right)_S=\left(1+\frac{{\cal E}}{P}\right)\left(\frac{v_s}{c}\right)_S^2,
\label{gamma-1}
\end{equation}
where  derivation is performing at constant entropy $S$. Moreover,  $(v_s/c)_S=\sqrt{(\partial  P/\partial  {\cal E})_S}$ is the speed of sound in units of speed of light. The speed of sound  is an important  quantity  related directly with  the stiffness of the equation of state and play dramatic role on the maximum mass configurations.  In general,  since the adiabatic index is  a function of the baryon density, exhibits  radial dependence and consequently, provides local information for each neutron star configuration. Its values  vary from 2 to 4 in most  of the neutron stars equations of state ~\cite{Haensel-07}. In the specific case of a polytropic equation of state the adiabatic index   is a constant. The effective adiabatic indices, $\langle \gamma \rangle $ and $\gamma_{cr}$, in distinction to $\gamma$ (Eq.~(\ref{gamma-1}))  have a global character.
Both of them are directly related with the neutron star equation of state as well as with the strength of the gravitational field (see also Refs.~\cite{Bludman-73a,Bludman-73b,Ipser-70,Merafina-89,Negi-2001,Moustakidis-017,Negi-1999,Herrera-89,Herrera-94,Yousaf-2015,Yousaf-2016}).

Chandrasekhar, using the Schwarzschild constant-density interior solution (see below more details about this analytical solution),  found that in the Newtonian limit, the stability ensured when~\cite{Chandrasekhar-64}
\begin{equation}
\langle \gamma \rangle\geq \gamma_{cr}= \frac{4}{3}+\frac{19}{42}2\beta.
\label{gamma-Chns}
\end{equation}
Chandrasekhar,  employed  the approximation  that the adiabatic index $\gamma$ is a constant through the star~\cite{Chandrasekhar-64}. In particular, this approximation directly  relates the equation of state, which characterizes the fluid, with a possible stable configuration. In addition, Chandrasekhar~\cite{Chandrasekhar-64}, in the framework of the post-Newtonian approximation  using relativistic polytropes found the  relation
\begin{equation}
\gamma_{\rm cr}=\frac{4}{3}+C\left(\frac{P_c}{{\cal E}_c}\right),
\label{g-Chand-2}
\end{equation}
where $C=1.8095, 2.2615, 2.4968, 2.6325$ corresponds to  the polytropic index  $n=0,1,2,3$ respectively  and  $P_c$,  ${\cal E}_c$ are the central values of pressure and energy density. It should be noted that the ratio $P_c/{\cal E}_c$, can also be mentioned as a relativistic index and closely related with the compactness $\beta$ (see the extended discussion in section 5). Similar results have been found also by Tooper in a series of papers~\cite{Tooper-64,Tooper-65}. Moreover, Bludman~\cite{Bludman-73a,Bludman-73b} studied the stability of general relativistic polytropes and provided the formulae
\begin{equation}
\gamma_{\rm cr}\simeq \frac{4}{3}+1.73\left(\frac{P_c}{{\cal E}_c}\right)-0.31 \left(\frac{P_c}{{\cal E}_c}\right)^2.
\label{Bludm-1}
\end{equation}
It is worth to extent all these previous studies in order to examine  the dependence of $\gamma_{\rm cr}$ on the compactness parameter $\beta_{\rm max}$ (as well as  on the ratio $P_c/{\cal E}_c$)  close to the instability limit, which corresponds to the maximum mass configuration. Although, the study concerning the Newtonian or post-Newtonian case is universal, meaning that for low vales of $\beta$  ($\beta\ll 1$) the dependence of $\gamma_{\rm cr}$ is almost insensitive on the details of the EoS, this is not the case for high values of $\beta$. In this case the structure of a neutron star and the corresponding values of $\gamma_{\rm cr}$ are very sensitive  on the EoS. Since, especially for high values of densities, the uncertainty on pressure-energy dependence is appreciably, we expect  an influence on the values of $\gamma_{\rm cr}$.  In view of the above, we conclude that possible constraints on $\beta_{\rm max}$ may impose constraints on the high density behavior of the neutron star equations of state .

We can also study the stability of the equilibrium configuration by using the general properties of the central density as well as of the mass-radius relation~\cite{Weinberg-72}. In this case, the configuration is stable when the inequality $dM/d{\cal E}_c>0$  holds.  Actually,  this condition, due to its simplicity,  has been used extensively in the literature. However, it needs to be noted that this condition is just necessary and not sufficient and consequently,  it is weak compared to the criterion  $(\ref{ineq-gamma})$.

Now we will briefly discuss  four analytical solutions  of the TOV equations. In the case  of the  Schwarzschild constant-density interior solution (here after Uniform), the density is constant throughout the star
and the energy density and pressure read as~\cite{Weinberg-72,Schutz-85}
\begin{equation}
{\cal E}={\cal E}_c=\frac{3Mc^2}{4\pi R^3},
\label{Unif-E}
\end{equation}
\begin{equation}
\frac{P(x)}{{\cal E}_c}=\frac{\sqrt{1-2\beta}-\sqrt{1-2\beta x^2}}{\sqrt{1-2\beta x^2}-3\sqrt{1-2\beta }}, \quad x=r/R.
\label{Unif-Pr}
\end{equation}
This solution, although is far from being realistic,  has been applied extensively in the literature due to its simplicity.

The Tolman VII solution has been extensively employed to neutron star studies. Actually, its  physical realization has been studied  in detail in Ref.~\cite{Raghoonundun-15}. In this case, the energy density and the pressure read as (for more details see~\cite{Lattimer-2001,Moustakidis-017})
\begin{equation}
\frac{{\cal E}(x)}{{\cal E}_c}=(1-x^2), \quad {\cal E}_c=\frac{15Mc^2}{8\pi R^3},
\label{Tolm-E}
\end{equation}
\begin{equation}
\frac{P(x)}{{\cal E}_c}=\frac{2}{15}\sqrt{\frac{3e^{-\lambda}}{\beta }}\tan\phi-\frac{1}{3}+\frac{x^2}{5}.
\label{Tolm-Pr}
\end{equation}
In this solution the causality ensured for $\beta < 0.2698$. However, useful information and predictions are taken when applied even for higher values of $\beta$ (see for example Ref.~\cite{Sotani-18,Lattimer-05a}).

In the case of the  Buchdahl's solution the equation of state read~\cite{Buchdal-59,Buchdal-67}
\begin{equation}
{\cal E}(P)=12\sqrt{P^{*}P}-5P.
\label{EOS-Buch}
\end{equation}
It is worthwhile to notice that the Buchdahl's solution is applicable only for low values of the compactness ($\beta \leq 0.2$) since for higher values the speed of sound becomes infinite. However, its use helps to support the findings of the rest solutions even for low values of the compactness e.g. it forms a  {\it bridge} which connects the Newtonian and post-Newtonian limit with the relativistic one~\cite{Lattimer-2001,Moustakidis-017,Moustakidis-016}.

The Nariai IV solution~\cite{Nariai-50,Nariai-51,Nariai-99}, although is very complicated, it provides useful insights because is one of the physically interest solutions. In this case, the energy density and pressure are complicated functional of the parametric variable $r'$, which is related with the distance $r$ (for the definitions of the involved functions and constants and for more details see Ref.~\cite{Moustakidis-017}).

All these solutions have the required  property  that the derived  density and pressure   vanish at the surface of the star (except of the Schwarzschild constant-density interior solution). In general, the selected solutions exhibit realistic behaviour and can be used as a guide  to establish  some universal approximations. In particular, while the unrealistic Uniform solution has been used  by Chandrasekhar~\cite{Chandrasekhar-64} in order to prove his famous expression~(\ref{gamma-Chns}),  its main drawback is the infinite value of the speed of sound.  In the case of Tolman VII solution, the causality ensured for $\beta < 0.2698$. However, useful information and predictions are taken when applied even for higher values of $\beta$. Thus, Lattimer and Prakash~\cite{Lattimer-05a} have demonstrated, using the Tolman VII solution, that the largest measured mass of a neutron star establishes an upper bound to the energy density of observable cold matter. Moreover, while in the Nariai IV solution the causality ensured for $\beta < 0.2277$, its extension for higher values was applied successfully~\cite{Moustakidis-017,Lattimer-05a,Moustakidis-016}.

\section{Maximum mass and maximum rotation frequency}
It is known that rotation increases the maximum mass ($M_{\rm max}^{\rm stat}$) of a corresponding stationary neutron star. In this case, we face two extreme configurations: a)  with maximum mass $M_{\rm max}^{\rm rot}$ and b) with  maximum rotation frequency $f_{\rm max}$ (known as Kepler frequency)~\cite{Haensel-07}. These configurations do not coincide but since are very close to each other (with  high accuracy)  we do not distinguish them.   Moreover, it was found that the maximum frequency can be expressed, with  high accuracy, in terms of mass and radius of the non-rotating configuration with the maximum mass (see Ref.~\cite{Haensel-07} and references therein). A precise formulae which relate $M_{\rm max}^{\rm stat}$ with the maximum mass and the compactness parameter $\beta_{\rm max}^{\rm stat}$ of the the static maximum-mass configuration is found by Haensel et. al.~\cite{Lasota-96, Haensel-99,Haensel-16}
\begin{equation}
f_{\rm max}\simeq 15.125\  \beta_{\rm max}^{3/2} (1+1.6164\beta_{\rm max})\left(\frac{M_{\odot}}{M_{\rm max}^{\rm stat}}\right)\ {\rm kHz}.
\label{F-max-1}
\end{equation}
It is worth to point out the strong dependence of $f_{\rm max}$ on $\beta_{\max}^{\rm stat}$ and consequently, via the adiabatic index, on the high density dependence area of the EoS. The above expression can be used to constrain an absolute lower bound of the maximum frequency of rigid rotation (for example by measuring the upper bound on the surface red-shift  of a  non-rotating neutron star) and consequently to impose useful constraints on the EoS and vise-versa.

\section{Results and Discussion}
We employ a large number of published realistic equations of state for neutron star matter based on various theoretical nuclear models. We calculate  both the effective averaged and the critical adiabatic indices for each configuration and mainly focus on the adiabatic indices corresponding to  the maximum mass configuration. The calculation recipe is the following: Firstly, we solve the TOV equations in order to determine the M-R dependence as well as the corresponding energy density and pressure configurations. Mainly, we are interested for the  maximum mass, the corresponding radius, the ratio $P_c/{\cal E}_c$ and the corresponding compactness $\beta$  for each  case.
Secondly, for each configuration we determine $\langle \gamma \rangle$ and $\gamma_{\rm cr}$. The onset of instability is found from the equality $\langle \gamma \rangle=\gamma_{\rm cr}$. The corresponding compactness parameter, denoting as  $\beta_{\rm max}$.

There is also a second criterion, which defines the stability limit according to the  equality  $dM/d{\cal E}_c=0$, providing an additional value of $\beta$ for the maximum mass configuration. Now, in general, since  $\langle \gamma \rangle$ and $\gamma_{\rm cr}$ are functionals of the trial function $\xi(r)$, we expect that the calculated values of $\beta$, for the two methods,  will not coincide. In these cases, we will consider as the most optimum trial function $\xi(r)$  the one that produces values of $\beta$, as close as possible, to the second method. In particular, we found that the trial function (\ref{xi-1}) (indicated as TF-1) is the optimal one, leading to an  error, in the most of the cases,  less than $1\%$.

In Fig.~1  the radius-mass relation  is drawn  using the selected EoSs. One can see that the majority of the EoSs reproduce the recent observation of two-solar mass neutron stars. It is obvious that the various predictions cover a wide range of the maximum neutron star masses and the corresponding radii.

In Fig.~2 we display  the dependence of  $\gamma_{\rm cr}$   as a function of the compactness parameter $\beta$ for all the employed EoSs by using the optimal trial function~(\ref{xi-1}). In particular, for the trial function~(\ref{xi-3})  we use the parametrization  $a_1=1/10R^2$, $a_2=1/5R^4$ and $a_3=3/10R^6$. The results of the four analytical solutions have been also included for comparison. The blue dots correspond to all configurations with neutrally stable equilibrium as results of the equality $\langle \gamma \rangle=\gamma_{\rm cr}$. These configurations correspond to the one with the largest possible central density reachable for stable configuration of a given mass. In the case of the Tolman VII solution, the results using  the  trial function TF-1~(\ref{xi-1}) have been also  included. In this case, the onset of instability is indicated  by the red star and corresponds to $\beta=0.3475$ and $\gamma_{\rm cr}=3.85$.
It is remarkable that the use of the Tolman VII solution, leads to results very close to the predictions by using realistic equations of state. The other two  analytical solutions (Buchdahl's and Nariai IV) lead to stable configuration in each case~(\cite{Moustakidis-017}). The Uniform solution is always used as a guide for stable configuration mainly for low values of the compactness $\beta$ (see expression (\ref{gamma-Chns})).

The most distinctive feature in Fig.~2 is the remarkable  unanimity of all equations of state and consequently the   occurrence of a model-independent  relation  between $\gamma_{\rm cr}$ and $\beta_{\rm max}$, at least for any stable configuration. The above finding, clearly expected for low values of the compactness $\beta$ (since all equations of state converge for low values of density). However, at high densities of the equations of state, where there is a considerable uncertainty,   this result was not obvious.
In any case, as a  consequence of the convergence, both for low and high values of the compactness the majority of the points indicate the onset of the instability, located in the mentioned trajectory. In particular, we found that  the simple expression
\begin{equation}
\gamma_{\rm cr}(\beta)=y_0+A_1e^{\beta/t_1}
\label{gr-beta-fit-anal}
\end{equation}
reproduces very well the numerical results  due to the use of realistic equations of state. Equation (\ref{gr-beta-fit-anal}) is the relativistic expression for the critical value of the adiabatic index and can be considered as the relativistic generalization of the post-Newtonian approximation~(\ref{gamma-Chns}). The parametrization is provided in Table~1.

The results of the analytical solutions, in each case, can be parameterized according to the expression (see details in Table~1)
\begin{equation}
\gamma_{\rm cr}(\beta)=y_0+A_1e^{\beta/t_1}+A_2e^{\beta/t_2}.
\label{gr-beta-fit-anal-2}
\end{equation}
Obviously, there is a small deviation between the results of the realistic equations of state and the analytical solutions Tolman VII, Nariai IV and Buchdahl. It is worth to notice that the Tolman VII solution reproduces very well the numerical results, especially for high values of the compactness.  In general, the analytical solutions lead to lower values of the adiabatic index $\gamma_{\rm cr}$, compared to the realistic EoS. In particular, the Uniform solution provides the lower limit for $\gamma_{\rm cr}$, especially for high values of the compactness and close to the instability limit.  However, the general trend is similar and useful insight can be gained concerning the reliability of analytical solutions. The stable configurations, independently of the equation of state, correspond to a universal relation between $\gamma_{\rm cr}$ and  $\beta$. One can safely conclude that $\gamma_{\rm cr}$ is an intrinsic property of  neutron stars (likewise the parameter $\beta$) which reflects the relativistic effects on their structure. In particular, $\gamma_{\rm cr}$ exhibits  a linear dependence  with $\beta$ in the Newtonian and post-Newtonian regime but a more complicated behavior in the relativistic regime (see also the Appendix).

Actually, the above finding may help to impose constraints to the equation of state of neutron star matter. For example, the accurate and simultaneously observation of possible maximum neutron star mass and the corresponding radius will constrain the maximum values of the compactness and consequently the maximum value of the  adiabatic index $\gamma_{\rm cr}$. In any case, useful insights may be gained by the use of the expression (\ref{gr-beta-fit-anal}) with the parametrization given in Table 1 (Realistic EoS).

In order to clarify further the effects of the trial functions $\xi(r)$ on the results, we present the Fig.~3. In particular, in Fig.~3 we display the dependence of the critical adiabatic index, $\gamma_{\rm cr}$, which corresponds to the onset of instability ($\gamma_{\rm cr}$=$\langle \gamma \rangle$ at this point), as a function of the compactness parameter $\beta_{\rm max}$ using the selected trial functions (\ref{xi-1}), (\ref{xi-2}), (\ref{xi-3}) and (\ref{xi-4}). The most distinctive feature in this case, is the occurrence of an almost linear dependence (in the region under study, e.g. on the maximum mass configuration) between the adiabatic index and the compactness $\beta_{\rm max}$.  Obviously, the use of the trial function $\xi(r)$ affects mainly the values of $\gamma_{\rm cr}$ (for the same $\beta_{\rm max}$) but not the linear dependence.

Moreover, in Fig.~4   we display the  $\gamma_{\rm cr}$, as a function of the compactness parameter $\beta$, for the selected EoSs, using  the trial function TF-1~(\ref{xi-1}) and  the optimal trial function (OTF) in each EoS, which corresponds to the one with the smallest error. The expression~(\ref{gr-beta-fit-anal}) which reproduces the numerical results corresponding to the trial function (\ref{xi-1}) is also included. Obviously, using the optimal trial function in each EoS the rearrangement of the results becomes more ordering. However, the deviation of using the trial function TF-1 (which is the optimal one in the  most of the cases) is negligible.

In Fig.5(a) we display the dependence of $\gamma_{\rm cr}$ on the ratio $P_c/{\cal E}_c$ (which corresponds to the maximum mass configuration). The symbols correspond to the results originated from the use of realistic equations of state while the results of the four analytical solutions have been also included for comparison. In general, in the case of realistic equation of state, $\gamma_{\rm cr}$ is an increasing function of the ratio $P_c/{\cal E}_c$ without obeying in a specific formulae. However, we found that the expression
\begin{equation}
\gamma_{\rm cr}=\gamma_0+C_1\left(\frac{P_c}{{\cal E}_c} \right)+C_2\left(\frac{P_c}{{\cal E}_c} \right)^2
\label{gr-epc-fit-anal}
\end{equation}
reproduces very well the numerical results of the analytical solutions. The parameters $\gamma_0, C_1,C_2$ are displayed  in Table~1.  In Fig.~5(b) displayed the dependence of $\gamma_{\rm cr}$ on $M_{\rm max}$. In Fig.~5(c) we plot   $\gamma_{\rm cr}$ as a function of the radius corresponding to the maximum mass configuration, $R_{\max}$. Obviously, in these cases, the dependence is almost random  and consequently  is unlikely to impose constraints from these kind of correlations.

It is known that for low values of $\beta$ (in the framework of Newtonian and post-Newtonian approximation) there is a very simple and universal  linear correlation between $\beta$ and the ratio $P_c/{\cal E}_c$. In particular, in the case of the analytical solutions of the TOV equations (Uniform, Tolman VII, Buchdahl's and Nariai IV) we get in each case, by employing a Taylor expansion, the approximated simple  relation
\begin{equation}
\frac{P_c}{{\cal E}_c} \simeq\frac{\beta}{2}.
\label{PcEc-b}
\end{equation}
Moreover, in the case of the Newtonian limit e.g. using the Lane-Emden equation with the  polytropic  equation of state $P=K({\cal E}/c^2)^{\Gamma}
=K({\cal E}/c^2)^{1+\frac{1}{n}}$
that is
\begin{equation}
\frac{1}{\xi^2}\frac{d}{d\xi}\xi^2\frac{d\theta}{d \xi}=-\theta^n
\label{L-E-1}
\end{equation}
with $\theta(\xi_0)=0$,
we get for the total mass and radius~\cite{Shapiro-83}
\begin{equation}
M=4\pi\left[\frac{(n+1)K}{4\pi G}  \right]^{3/2} \left(\frac{{\cal E}_c}{c}\right)^{(3-n)/2n}\xi_{0}^2|\theta'(\xi_0)|
\label{Pol-M}
\end{equation}
and
\begin{equation}
R=\left[\frac{(n+1)K}{4\pi G}   \right]^{1/2}\left(\frac{{\cal E}_c}{c}\right)^{(1-n)/2n}\xi_{0}.
\label{Pol-R}
\end{equation}
Combining Eqs.~(\ref{Pol-M}) and (\ref{Pol-R})
we found
\begin{equation}
\frac{P_c}{{\cal E}_c} =\frac{\beta}{2}\left(\frac{n+1}{2}\xi_0|\theta'(\xi_0)|  \right)^{-1},
\label{Poly}
\end{equation}
or in general
\begin{equation}
\frac{P_c}{{\cal E}_c}=\frac{\beta}{2}{\cal F}(\xi_0,n),
\label{F-1}
\end{equation}
where ${\cal F}(\xi_0,n)$ is a function of $\xi_0$ and the polytropic index $n$.
More precisely, we found that for $n=0, 0.5, 1, 1.5, 2, 3, 4$  (correspondingly $\Gamma=\infty, 3, 2, 5/3, 3/2, 4/3, 5/4$) the respectively function is ${\cal F}(\xi_0,n)=$ $1$, $0.97$, $1$, $1.077$, $1.204$, $1.709$, $3.332$. Concluding, for $0<n<2$  we get ${\cal F}(\xi_0,n)\simeq 1$.

Since it is worth to examine this dependence in the relativistic limit, we display in Fig.~5(d) the dependence of $P_c/{\cal E}_c$ on the compactness $\beta_{\rm max}$. Firstly, we can see that symbols originated from the use of realistic EoSs obey to a general  trend. A similar trend is obtained by employing the analytical solutions. In particular, the Tolman VII and Nariai IV solutions reproduce very well the results of realistic calculations. Consequently, the Tolman VII solution may by used as a guide for an almost universal dependence between $P_c/{\cal E}_c$ and  $\beta_{\rm max}$ that is in the critical point  between stable and unstable  configuration. Moreover, this correlation may help to constrain the maximum value of the ratio  $P_c/{\cal E}_c$ and consequently, the maximum density in the universe by the help of accurate measurements of the maximum value of the compactness.

To be more specific, from recent observations of the GW170917 binary system  merger, Bauswein {\it et al}.,~\cite{Bauswein-17} propose a method to constrain some neutron stars properties. In particular, they found that the radius $R_{\rm max}$ of the nonrotating maximum-mass configuration must be larger than $9.6_{-0.04}^{+0.15} \ {\rm km}$. Almost simultaneously,  Margalit and Metzger~\cite{Margalit-17} combining electromagnetic and gravitational-wave information on the binary neutron star merger GW170817, constrain the upper limit of $M_{\rm max}$ according to $M_{\rm max}\leq 2.17 M_{\odot}$.  The combination of the two suggestions leads to an absolute  maximum value of compactness, which is equal to $\beta_{\rm max}=0.333_{-0.005}^{+0.001}$. The use of this value with the help of the Fig.~2 and  5(d) will impose constraints both on the maximum values of the index  $\gamma_{\rm cr}$  and the ratio $P_c/{\cal E}_c$. According to expression (\ref{gr-beta-fit-anal}), constraint on the $\gamma_{\rm cr}$ can be imposed, which is $\gamma_{\rm cr,max}=3.381_{-0.095}^{+0.020}$, correspondingly to $\beta_{\rm max}$. Even more, a large number of realistic equations of state must be excluded. Some previous and recent efforts, to constrain the compactness of neutron stars, have been provided in Refs.~\cite{Miller-98,Hambaryan-17,Rosso-17,Chen-15,Alsing-18,Ravenhall-94}.

In fig.~6(a) we display the dependence of the maximum rotating frequency on the critical adiabatic index (it is wort to indicate, in order to avoid any confusion, that in the present study  $M_{\rm max}^{\rm stat}$,  $R_{\rm max}^{\rm stat}$ and $\beta_{\rm max}^{\rm stat}$  correspond  to $M_{\rm max}$, $R_{\rm max}$ and  $\beta_{\rm max}$ respectively). Obviously, while $f_{\rm max}$ is an increasing function of $\gamma_{\rm crit}$, the correlation is not so restrictive. However, the most important finding (see also the fig.~6(b)) is the derivation of an absolute  lower upper bound of the maximum rotation rate close to the value $1460 $ Hz. The observation of neutron stars rotating with a spin $f>1460$ Hz, will exclude a number of the selected EoSs. In fig.~6(b) we display also the dependence of  $f_{\rm max}$ on $\beta_{\rm max}^{\rm stat}$  while in fig.~6(c) the dependence of $f_{\rm max}$ on the mass which corresponds to the static maximum mass configuration is provided. In this case, the dependence is random. However, the dependence of  $f_{\rm max}$  on the radius which corresponds to the static maximum mass configuration, exhibits a more restrictive dependence. In particular, $f_{\rm max}$ is a decreasing function of $R_{\rm max}^{\rm stat}$ i.e. the maximum rotation rate is expected to be observed in low size neutron stars.

In any case,  further theoretical and  observational studies, as well as refined combinations of them, are necessary before accurate, reliable  and robust constraints to be inferred.

\begin{table}[h]
 \begin{center}
\caption{The parametrization of the analytical formulae  (\ref{gr-beta-fit-anal}),  (\ref{gr-beta-fit-anal-2}) and (\ref{gr-epc-fit-anal}) using realistic equations of state as well as four analytical solutions  (using  the trial function TF-1~(\ref{xi-1})). The case mentioned as {\it Realistic EoS}, reproduces the averaged results of the realistic equations of  state. } \label{t:1}
\vspace{0.5cm}
\begin{tabular}{|l|c|c|c|c|c|c|c|c|}
\hline
 Solution    &  $y_0$  & $A_1 $     & $t_1$   & $A_2$     & $t_2$    &$\gamma_0$ & $C_1$  &  $C_2$ \\
\hline
Realistic EoS       &  1.23333    & 0.10425   &   0.11007     &   &  &   &   &    \\
\hline
Tolman VII       &  1.18654    & 0.14938    & 0.15293       &  0.00011 & 0.03731 & 1.32055  & 2.45877  &  -0.36691  \\
\hline
Buchdahl     &   1.04258    & 0.28285    &  0.27558      &  0.00792  &  0.07695 &  1.33344     &   2.25592     & -1.28137    \\
\hline
Nariai IV     &  1.13470     &  0.20200   &  0.16781       &  0.00015  &  0.04016 & 1.33094    &  2.68839  &  -0.45055  \\
\hline
Uniform       &   1.18955    &  0.14587   &   0.17682     &  0.00009   & 0.04140  &  1.32743    &  1.94115      &-0.08660   \\
\hline
\end{tabular}
\end{center}
\label{ntptm14}
\end{table}

\begin{figure}
\centering
\includegraphics[height=8.1cm,width=9.1cm]{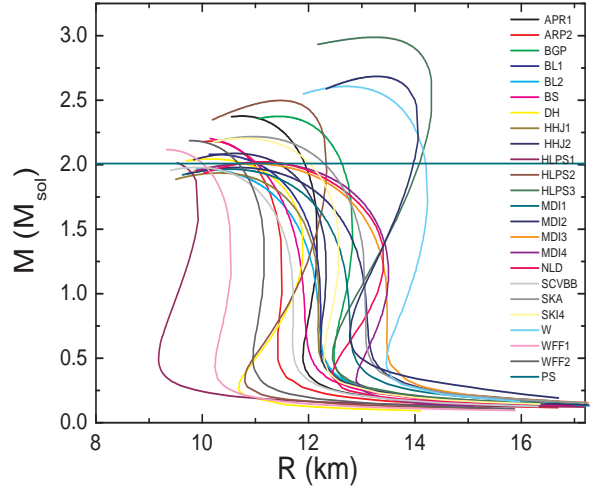}\
\caption{Mass-radius trajectories for the selected EoSs.  }
\label{}
\end{figure}
\begin{figure}
\centering
\includegraphics[height=9.1cm,width=10.1cm]{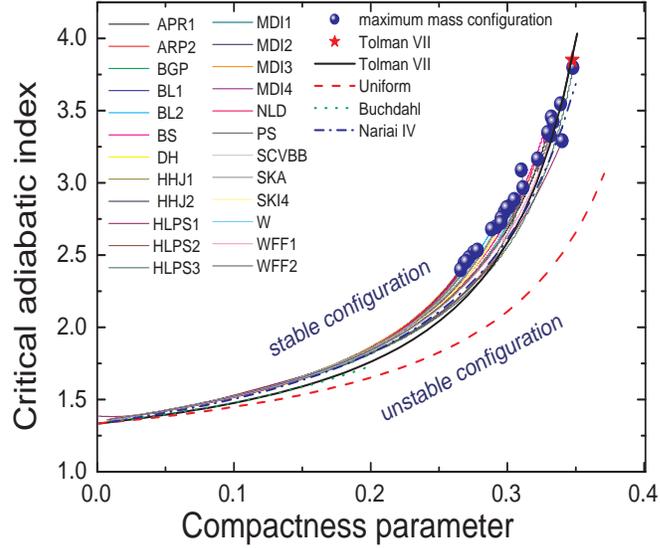}\
\caption{The critical adiabatic index, $\gamma_{\rm cr}$, as a function of the compactness parameter $\beta$, for the selected EoSs (using  the trial function TF-1~(\ref{xi-1})). The results of the four analytical solutions, using  for consistency the trial function TF-1~(\ref{xi-1}), have been also included for comparison (for more details see text). The blue dots correspond to the onset of instability as a result of the equality $\langle \gamma \rangle=\gamma_{\rm cr}$. The onset of instability for the Tolman VII solution is indicated by  the red star (for the TF-1).}
\label{}
\end{figure}
\begin{figure}
\centering
\includegraphics[height=9.1cm,width=10.1cm]{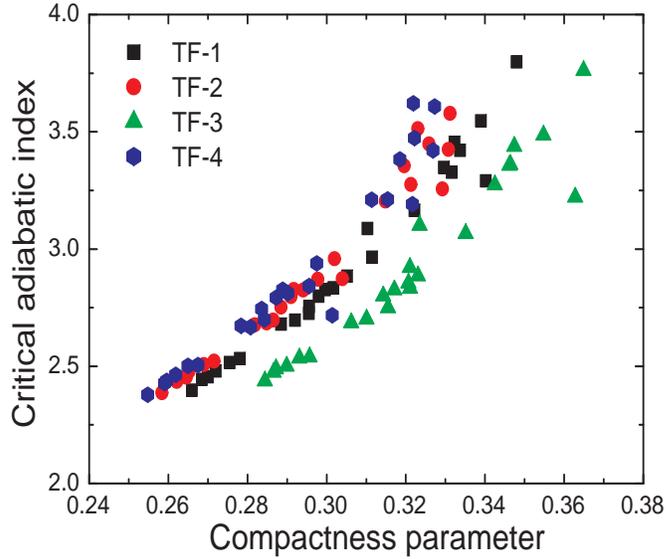}\
\caption{ The critical adiabatic index $\gamma_{\rm cr}$ as a function of the compactness parameter $\beta$ for the selected EoSs. The points correspond to the onset of instability for the four selected trial functions $\xi(r)$.   }
\label{}
\end{figure}
\begin{figure}
\centering
\includegraphics[height=8.1cm,width=9.1cm]{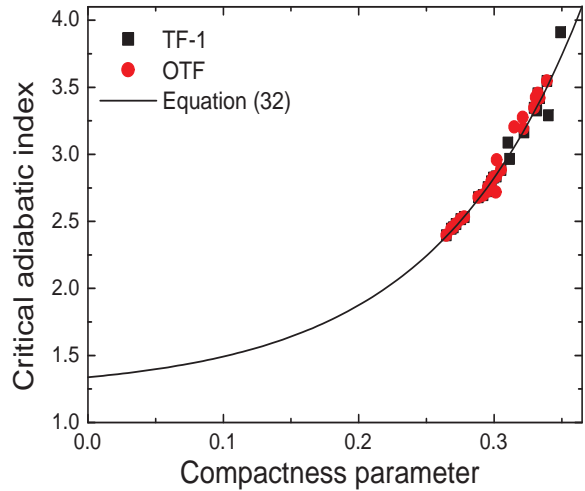}\
\caption{The critical adiabatic index, $\gamma_{\rm cr}$, as a function of the compactness parameter $\beta$, for the selected EoSs, using  the trial function TF-1~(\ref{xi-1}) (squares) and the optimal trial function (OTF) in each EoS (dots). The expression~(\ref{gr-beta-fit-anal}) (the parametrization is provided in Table~1) which reproduces the numerical results corresponding to the trial function (\ref{xi-1}) is also included.     }
\label{}
\end{figure}
\begin{figure}
\centering
\includegraphics[height=7.1cm,width=7.1cm]{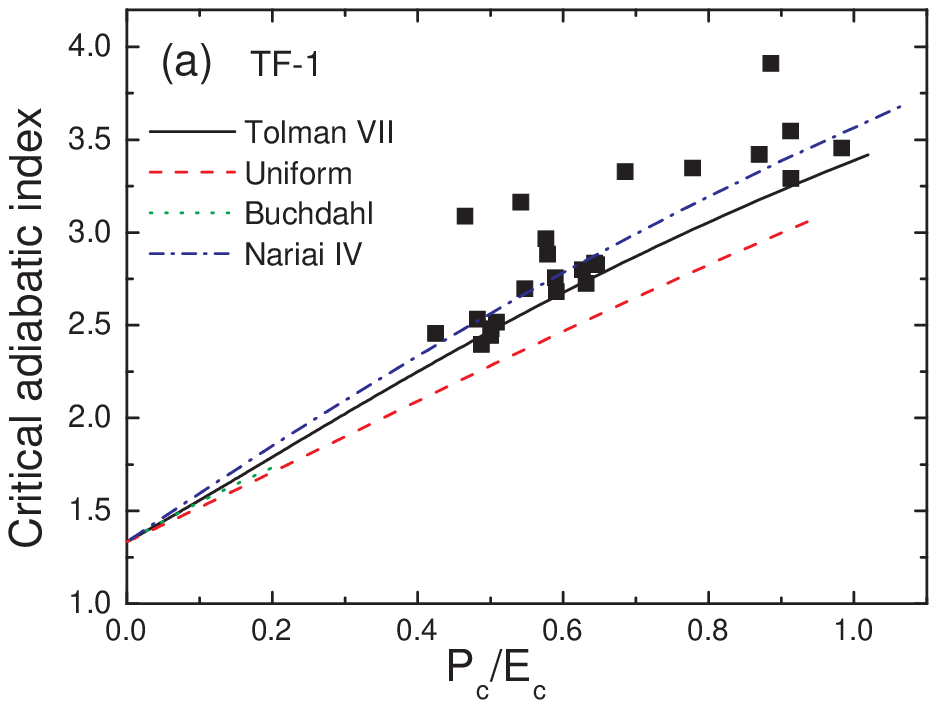}\
\includegraphics[height=7.1cm,width=7.1cm]{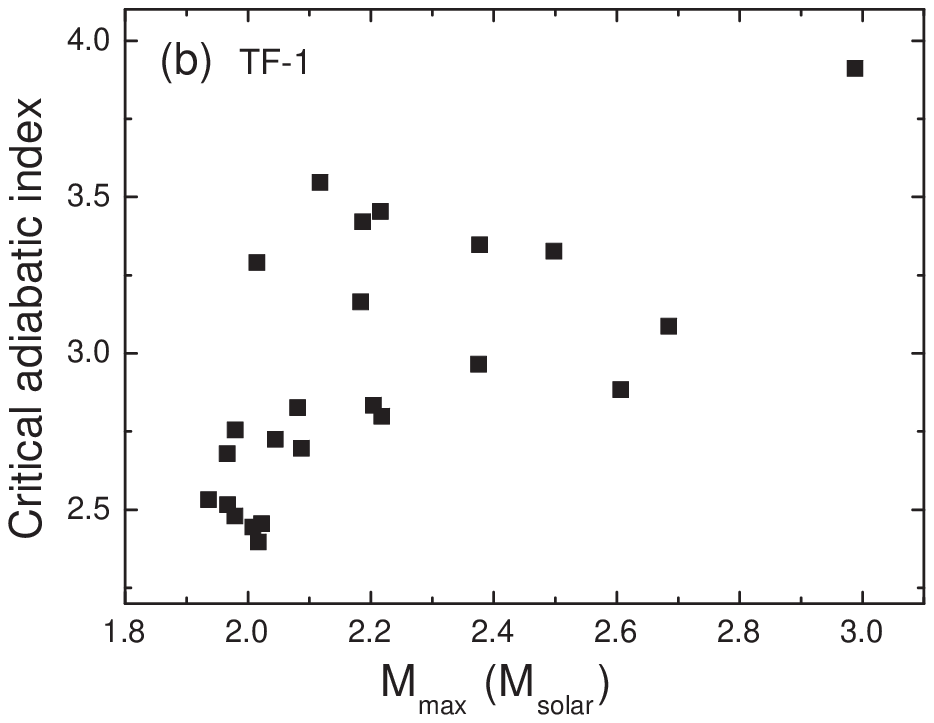}\\
\includegraphics[height=7.1cm,width=7.1cm]{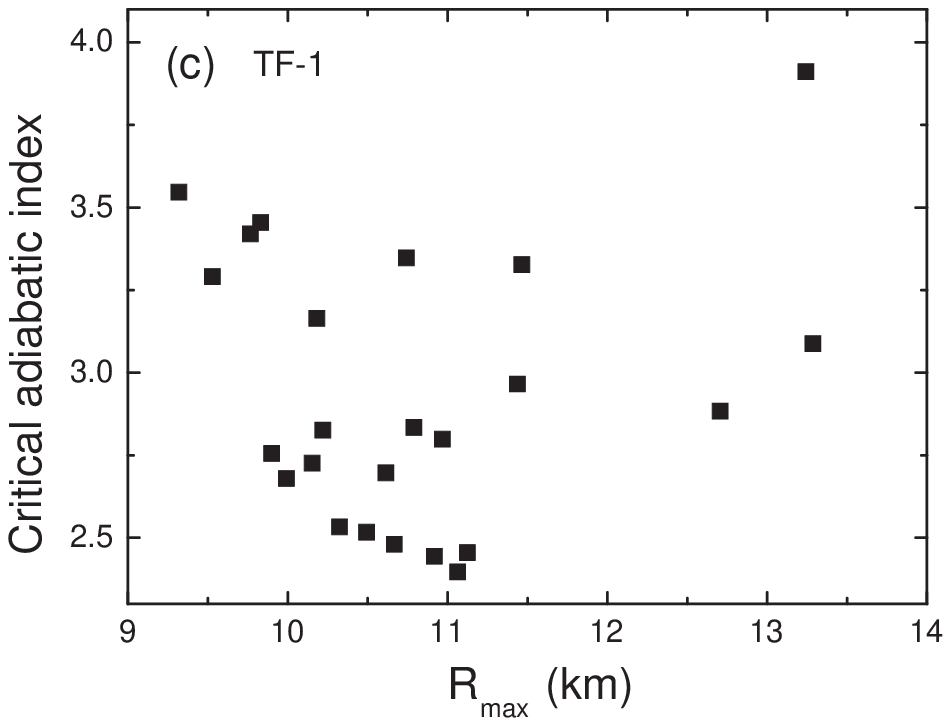}\
\includegraphics[height=7.1cm,width=7.1cm]{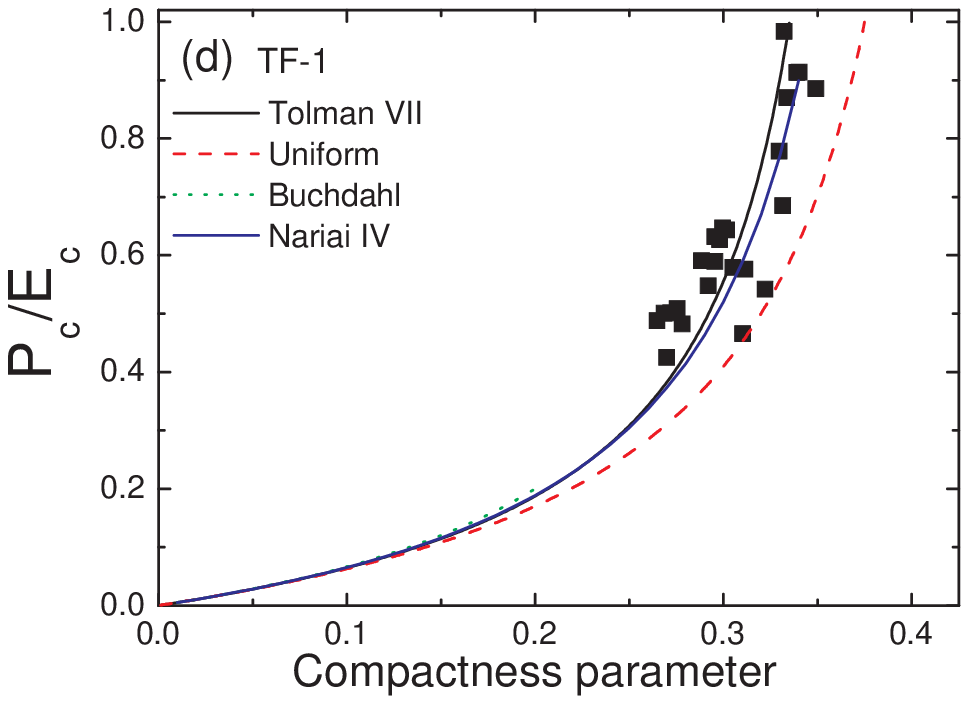}\
\caption{(a) The critical adiabatic index, $\gamma_{\rm cr}$, as a function of the ratio $P_c/{\cal E}_c$ for the selected EoSs (the dots correspond to the onset of instability in each case) and for the trial function TF-1 ~(\ref{xi-1}). The results of the four analytical solutions have been also included for comparison, (b) the  $\gamma_{\rm cr}$ as a function of the maximum mass, $M_{\rm max}$, for the EoSs, (c)  the  $\gamma_{\rm cr}$ as a function of the radius, $R_{\rm max}$, corresponding  to $M_{\rm max}$ for the selected EoSs, and (d) the ratio $P_c/{\cal E}_c$ as a function of the compactness parameter, $\beta$, for the selected EoSs while the results of the four analytical solutions have been also included for comparison.    }
\label{}
\end{figure}

\begin{figure}
\centering
\includegraphics[height=7.1cm,width=7.3cm]{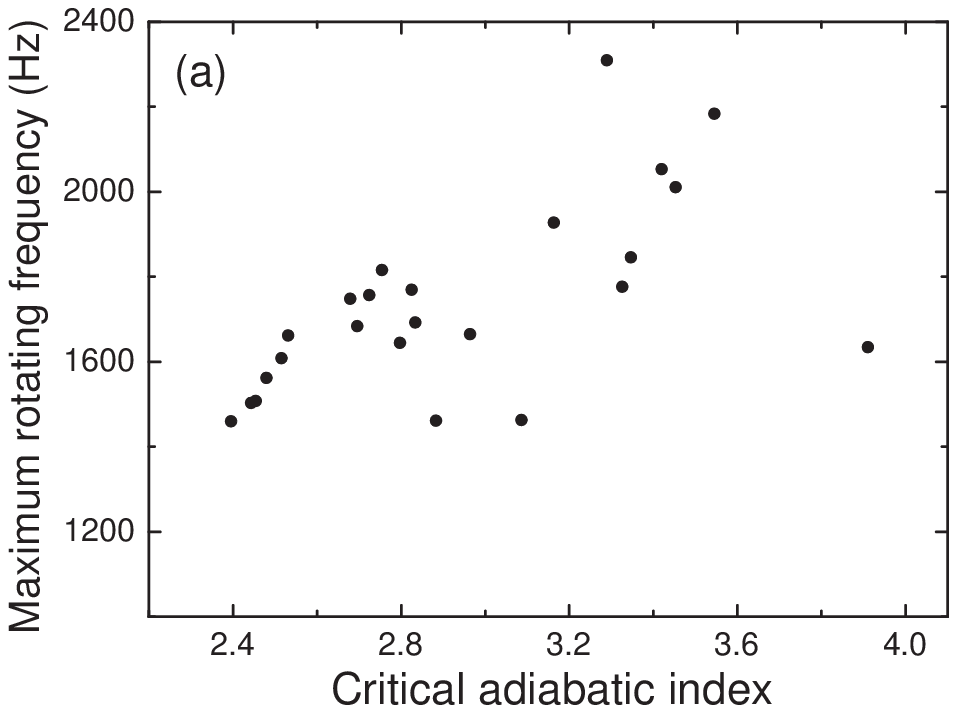}\
\includegraphics[height=7.1cm,width=7.3cm]{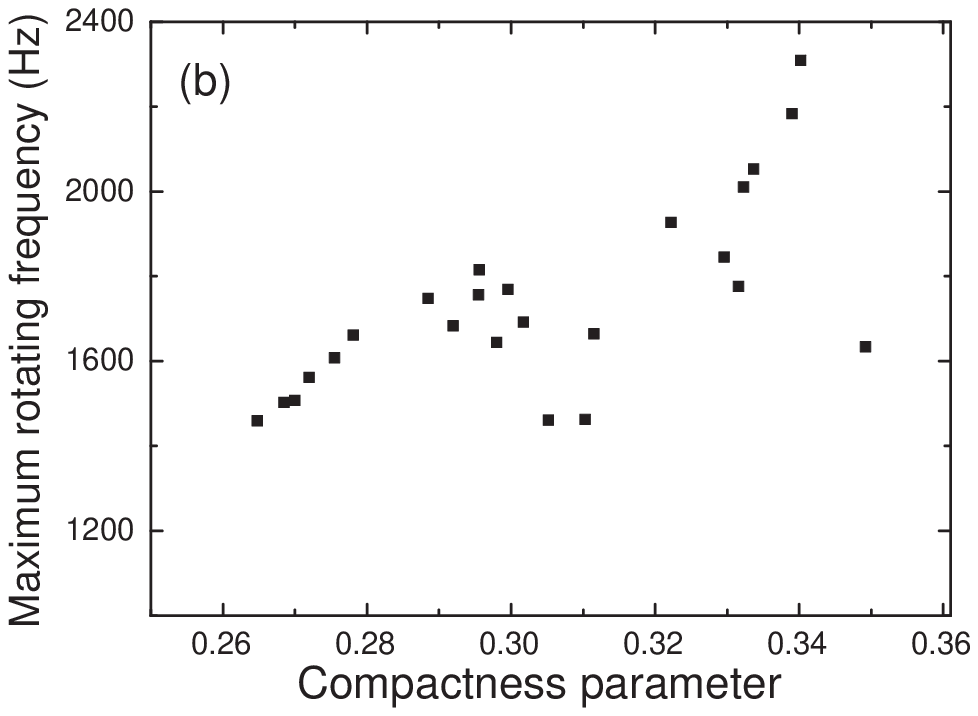}\\
\includegraphics[height=7.1cm,width=7.3cm]{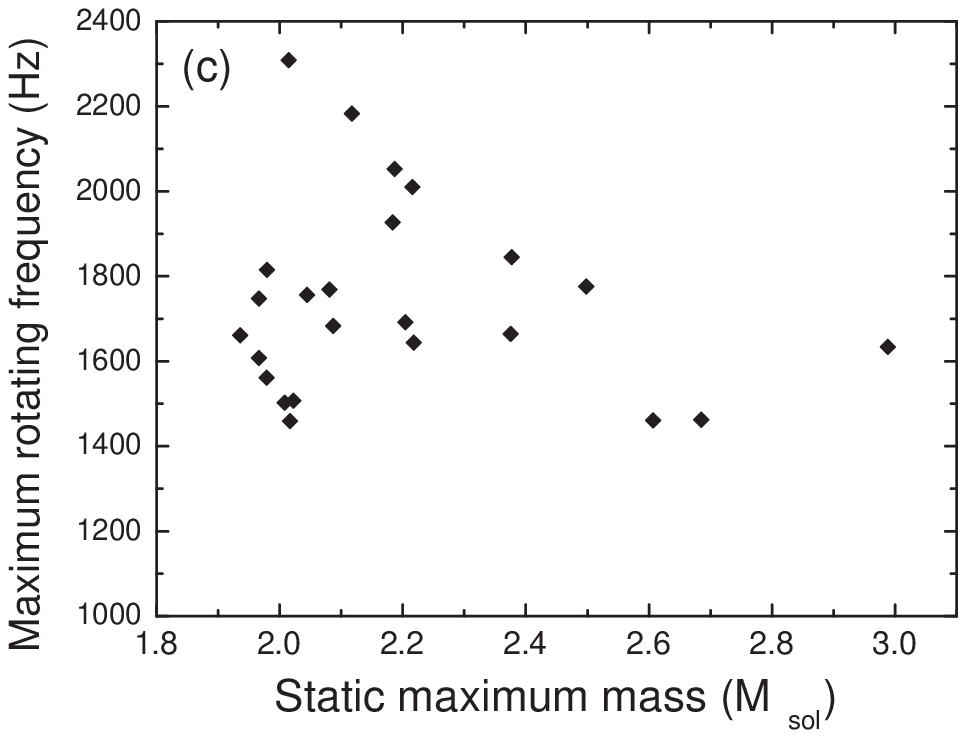}\
\includegraphics[height=7.1cm,width=7.3cm]{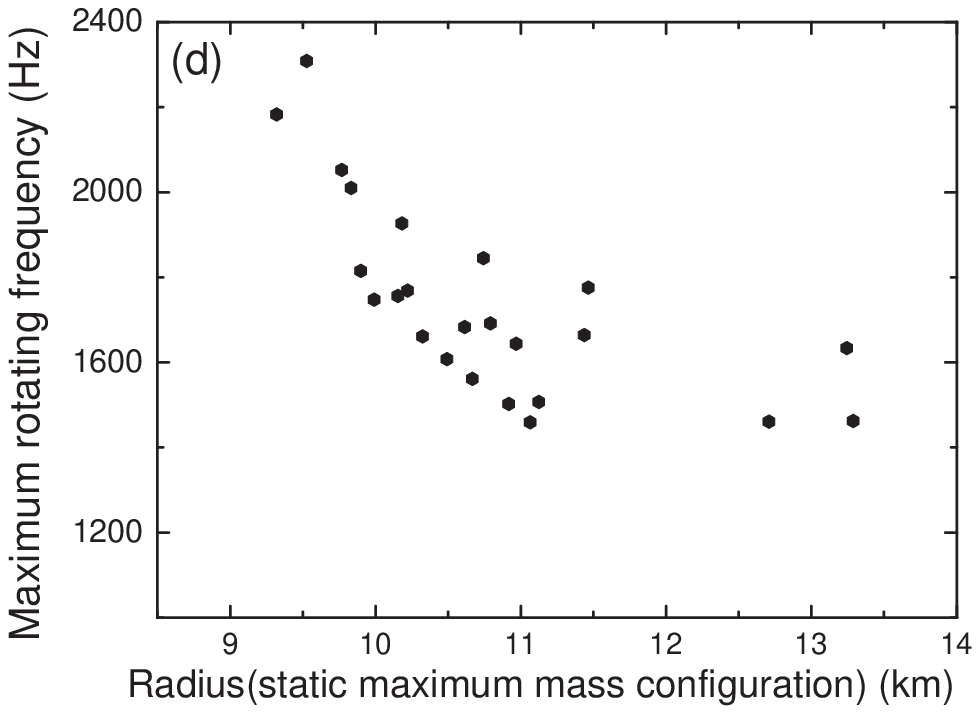}\
\caption{The maximum rotating frequency $f_{\rm max}$ for the selected EoSs,  as a function: (a) of the critical adiabatic index, $\gamma_{\rm cr}$, (b) of the compactness parameter $\beta_{\rm max}^{\rm stat}$ which corresponds to the static maximum mass configuration, (c) of the static maximum mass $M_{{\rm max}}^{\rm stat}$ and (d) of the radius $R_{\rm max}^{\rm stat}$ which corresponds to the static maximum mass configuration.     }
\label{}
\end{figure}

\section{Concluding remarks}
We suggested a new method  to constraint the neutron star equation of state by means of the stability condition introduced by Chandrasekhar~\cite{Chandrasekhar-64}. We found that the predicted critical adiabatic index, as  function of the compactness, for the most of the equations of state considered here (although they differ considerably at their maximum masses and  in how their masses are related to radii) satisfies a universal relation. In particular, the exploitation of  these results leads to a model-independent expression for the critical adiabatic index as a function of the compactness.      The expression (\ref{gr-beta-fit-anal}) (with the specific parametrization given in Table~1) reproduces very well this relation. The above finding may be added to the rest approximately EoS-independent relations~\cite{Ravenhall-94,Yagi-013a,Yagi-013b,Yagi-017,Yagi-014,Breu-016,Maselli-017,Silva-018,Silva-16}. These universal relations break degeneracies among astrophysical observations  and  leading to a variety of applications.  We  also found  that observations of high rotating neutron stars may help  to impose useful   constraints on the EoSs by using the dependence of maximum frequency on the compactness parameter corresponding to maximum mass configuration of a non-rotating neutron star and consequently on the adiabatic index (instability limit).
We state that additional theoretical and observational measurements of the bulk neutron star properties close to the maximum-mass configuration  will  help to impose robust constraints on the neutron star  equation of state or, at least, to minimize the numbers of the proposed EoSs.

\section*{Acknowledgments}
One of the authors (Ch.C.M)  would like to thank the Theoretical Astrophysics Department of the University of Tuebingen, where part
of this work was performed, for the warm hospitality and Professor K. Kokkotas  for his constructive comments and insights  during  the preparation of the manuscript. This work was partially supported by the  COST action PHAROS (CA16214) and the DAAD Germany-Greece grant ID 57340132.

\section{Appendix}
The numerical integration of the integrals related with the definition of $\langle \gamma \rangle$ and $\gamma_{\rm cr}$ can be easily  preformed.  However, following this procedure it is difficult to perceive  the final results. Actually, this is easy only in some approximated cases e.g. in  the Newtonian and post-Newtonian limit.  In the following, we try to generalize the finding of Chandrasekhar~\cite{Chandrasekhar-64} to even higher values of the compactness where the relativistic effects become important.
The expression of the critical adiabatic index, with the help of the TOV equations~(\ref{TOV-1}), (\ref{TOV-2}) and using the trial function $\xi(r)=r e^{\nu/2}$ (in order to be in consistent with the pioneering work of Chandrasekhar~\cite{Chandrasekhar-64}),  can also be written  as~\cite{Merafina-89}
\begin{eqnarray}
\gamma_{\rm cr}(\beta)&=&\frac{4}{3}+\frac{1}{36}\frac{\displaystyle \int_0^1 e^{(\lambda+3\nu)/2}\left[16P/{\cal E}_c+(e^{\lambda}-1)(P/{\cal E}_c+{\cal E}/{\cal E}_c)\right](e^{\lambda}-1) x^2  dx }
{\displaystyle \int_0^1 e^{(\lambda+3\nu)/2}(P/{\cal E}_c) x^2dx}
\nonumber \\
&+& {\cal C}_1(\beta)\frac{\displaystyle \int_0^1 e^{(3\lambda+3\nu)/2} \left[8P/{\cal E}_c+(e^{\lambda}+1)(P/{\cal E}_c+{\cal E}/{\cal E}_c)\right](P/{\cal E}_c) x^4  dx          }{\displaystyle \int_0^1 e^{(\lambda+3\nu)/2}(P/{\cal E}_c) x^2dx}
\nonumber \\
&+&{\cal C}_2(\beta)\frac{\displaystyle \int_0^1 e^{(5\lambda+3\nu)/2} (P/{\cal E}_c+{\cal E}/{\cal E}_c)(P/{\cal E}_c)^2 x^6  dx }{\displaystyle \int_0^1e^{(\lambda+3\nu)/2}(P/{\cal E}_c) x^2dx}, \quad x=r/R
\label{Gr-meraf}
\end{eqnarray}
where ${\cal C}_1(\beta)=\beta/3$ and ${\cal C}_2(\beta)=\beta^2$ for the Uniform solution and ${\cal C}_1(\beta)=5\beta/6$ and ${\cal C}_2(\beta)=25\beta^2/4$ for the Tolman VII solution. We performed  a Taylor expansion inside the integrals in each case and we found respectively for the Uniform and the Tolman VII solution that
\begin{equation}
\gamma_{\rm cr}(\beta)=\frac{4}{3}+\frac{38}{42}\beta\left(1+2.13\beta+4.65\beta^2+10.22\beta^3+{\cal O}(\beta^4)\right),
\label{Unif-taylor}
\end{equation}
\begin{equation}
\gamma_{\rm cr}(\beta)=\frac{4}{3}+\frac{38}{42}\beta\left(1.19+2.93\beta+7.34\beta^2+19.36\beta^3+{\cal O}(\beta^4)\right).
\label{Tolman-taylor}
\end{equation}
Obviously, the approximation~(\ref{Unif-taylor}), to a linear term,  confirms the Chandrasekhar expression~(\ref{gamma-Chns}).   The above expressions are good approximation for $\beta<0.2$. However, fail for higher values of $\beta$ and consequently  additional terms must be included. In particular,  $\gamma_{\rm cr}$ increases very fast for $\beta >0.25$ due to the strong effects of general relativity.


\end{document}